\documentclass[article,twocolumn,amsmath,floats,showpacs,nofootinbib]{revtex4}
\usepackage[usenames]{color}

\usepackage{graphicx}
\usepackage{epstopdf}
\usepackage{textcomp}
\usepackage{hyperref}
\usepackage{multirow}

\hypersetup{colorlinks=true}

\begin{document}

\title{PHONON SINGULARITIES ON VOLT-AMPERE CURVES OF NIOBIUM POINT CONTACTS}

\author{I.K. Yanson, N.L. Bobrov, L.F. Rybal'chenko, and V.V. Fisun}
\affiliation{B.I.~Verkin Institute for Low Temperature Physics and
Engineering, of the National Academy of Sciences
of Ukraine, prospekt Lenina, 47, Kharkov 61103, Ukraine and A. M. Gor'kov State University, Khar'kov
Email address: bobrov@ilt.kharkov.ua}
\published {\href{http://www.sciencedirect.com/science/article/pii/0038109884903193}{Solid State Communications}, Vol 50, No 6, 515, (1984)}
\date{\today}

\begin{abstract}The volt-ampere curves and their second derivatives were studied for niobium point contacts at low temperatures in the voltage range corresponding to the characteristic phonon energies. It was found that while for the dirty contacts in the normal state no PC spectra of phonons could be detected, in the superconducting state there were singularities in the I-V curves corresponding to maxima either in the first or in the second derivatives. The singularities observed were due to the energy dependence of the excess current. We suppose that the origin of these singularities is due to the inelastic transitions of electrons between chemical potentials of Cooper pairs at both sides of the contact, which differ in energy by $eV$. These transitions are possible if $\xi \left( 0 \right)\gtrsim d \sim{{\Lambda }_{\varepsilon }}$ ($\xi (0)$  being the coherence length, $d$, the contact diameter, ${{\Lambda }_{\varepsilon }}={{({{l}_{i}}\cdot {{l}_{\varepsilon }})}^{1/2}}$, where ${{l}_{i}}$  and ${{l}_{\varepsilon }}$  being the elastic and inelastic electron mean free paths, respectively).

\end{abstract}

\maketitle

It is known \cite{Yanson1}, that the point-contact spectroscopy (PCS) enables one to obtain direct information on the spectral function of the electron-phonon interaction (EPI) in metals being in normal (nonsuperconducting) state. The relevant theory \cite{Kulik} agrees well with the experiment for most of the metals studied \cite{Yanson2}, including superconductors with not too high critical fields Evidently, the PCS study of the high critical field superconductors (e.g., those of the A15 type having critical fields much higher than 100 kOe) is hardly possible now. Therefore, the question of how to extract the information about the EPI from the electrical characteristics of point contacts being in the superconducting state is of high present day importance.

Earlier there was an attempt to study the EPI in superconducting point contacts of tin \cite{Khotkevich}. It was found that the differential resistance curves possess maxima located in the voltage region of the characteristic phonon frequencies but no one to one correlation between the peak positions and phonon spectrum pecularities was observed. Because the volt-ampere curves studied are distinguished by the large nonlinearity and hysteresis along with the dramatic decrease of the excess current in the region of phonon energies the authors of \cite{Khotkevich}, were drawn to the conclusion about the thermal origin of the singularities observed. Then, in the theoretical paper \cite{Khlus1} the EPI in pure $S-c-S$ ($c$- constriction) contacts were shown to yield the energy dependence of the excess current at biases $eV>\Delta $ ($\Delta $  being the energy gap) leading to its decrease in the region of characteristic phonon frequencies. The nonlinearities due to this effect amount to only a small fraction of nonlinearity in the normal state and lead to the insignificant modification of the EPI spectrum although at definite conditions the phonon peaks should appear already in the first derivative.

Since among all the superconducting elements $Nb$ possesses very high critical parameters, one may expect that the point contacts made from this material appear to be most suitable for the discovering of the phonon singularities in their electrical characteristics in $S$-state, while for other superconductors they are masked by such unwanted effects as heating or destroyment of superconductivity by nonequilibrium energy distribution of electrons. The evidence of the smallness of these unwanted effects in high ohmic niobium contacts could be found, for example, in \cite{Divin}, where no noticeable decrease of the excess current was seen up to the biases corresponding to phonon energies.

In the present paper, we report on the experimental study of small nonlinearities of volt-ampere characteristics of niobium point contacts both in normal and superconducting states. We found that for the superconducting state in the dirty limit there were maxima either in the first or in the second derivatives of the I-V characteristics. The maxima were located at characteristic phonon energies and they disappeared after the transition to the normal state.

The small size metallic constrictions were created directly in liquid helium between two bulk niobium electrodes by the shear-type-deformation method \cite{Yanson2}. While moving the electrodes, one along another with the application of shear deformation, the metallic contact is created at the point where the isolation layer covering the surface of electrode is cracked. The remaining undestroyed isolation at the contact region serves as a support providing the contact with high electrical and mechanical stability The size of the contact is determined by the interelectrode pressure controlled from the outside of the cryostat. The electrodes with dimensions $2\times 2\times 12\text{ }m{{m}^{3}}$ were electrically sparked from the single crystal $Nb$ having the resistance ratio ${{\rho }_{300}}/{{\rho }_{res}}\approx 100$. We use two kinds of treatment to prepare the isolation layer on the surface of the electrodes. For the first group of electrodes the isolation layer was created by chemical etching in the mixture of acids $HF:HCl{{O}_{4}}:HN{{O}_{3}}$  taken in the equal volume ratios. However, the quality of the dielectric layer made by this treatment was not high enough In many cases the noticeable leakage currents through the supporting layer were observed. They hindered the registration of the true phonon nonlinearities. Much higher percentage of good quality junctions can be obtained for the contacts made from the electrodes treated as follows. Firstly they were heated by the sharp focussed electron beam gun ($U=4\div 5\,kV,\ I=5\div 10\ mA$) up to the near melting point during 5-7 minutes at the pressure of $(1\div 2)\times {{10}^{-7}}Torr$. Then, after cooling down to room temperature a few hundred Angstrom thickness $Al$ layer was vacuum deposited on the electrode facets which were subsequently oxidized in the boiling 30\%- water solution of $H_{2}O_{2}$. Measurements made on good contacts with both types of electrodes lead to the same results.

The volt-ampere characteristics and their derivatives were recorded by the standard modulation technique. For doing measurements in the normal state the superconductivity  was destroyed  by  magnetic field  up to 50 $kOe$ at the temperature of $4.2\ K$. Since switching of the magnetic field, both on and off, destroyed the contact mechanically, the measurements  for the same contact both in \emph{S-} and \emph{N-} states were carried out in helium vapor  after  temperature  was raised from 4.3 up to $10~K$.

\begin{figure}[]
\includegraphics[width=8cm,angle=0]{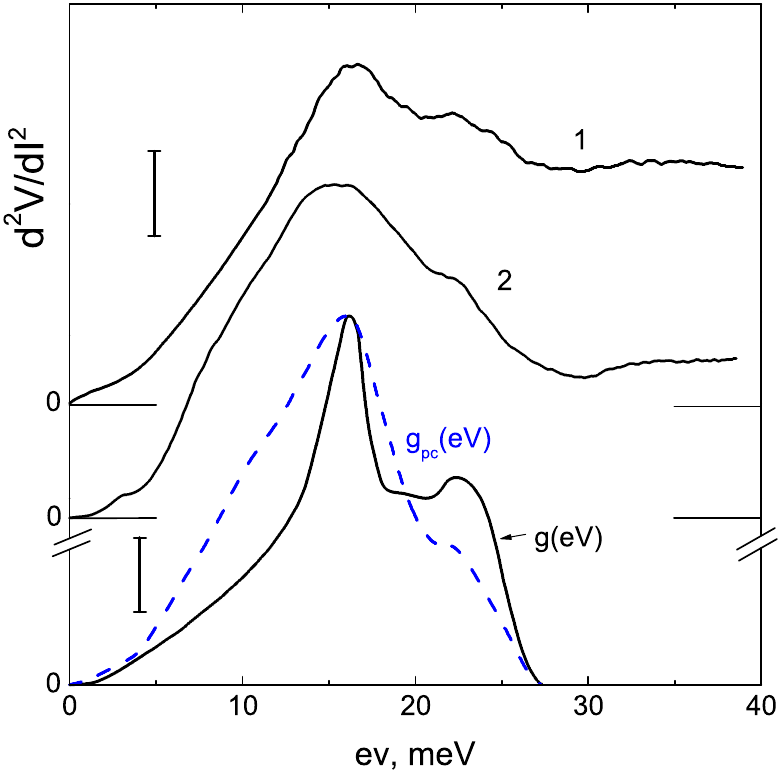}
\caption[]{Point-contact spectra of niobium in normal state (curves 1-2). The contact resistances, the modulation voltages at $V = 0$, and the length of the vertical segment with bars which calibrates the second harmonic signal ${{V}_{2}}\propto {{d}^{2}}V/d{{I}^{2}}$ are equal to 110 and $334\ \Omega $; 0.993 and $2.26\ mV$; 0.68 and $0.555\ \mu V$  for curves 1 and 2, respectively $T = 4.2\ K$, $H=40-50\ kOe$.  ${{g}_{pc}}\left( eV \right)$  - point- contact EPI function resulted from the spectrum 1 in the pure orifice model \cite{Yanson2}; $g(eV)$ is the tunneling EPI function taken from \cite{Robinson}. The vertical segment with bars is equal to 0.023 and 0.112 dimensionless units for $g_{pc}$ and $g$, respectively.}
\label{Fig1}
\end{figure}

\begin{figure}[]
\includegraphics[width=8cm,angle=0]{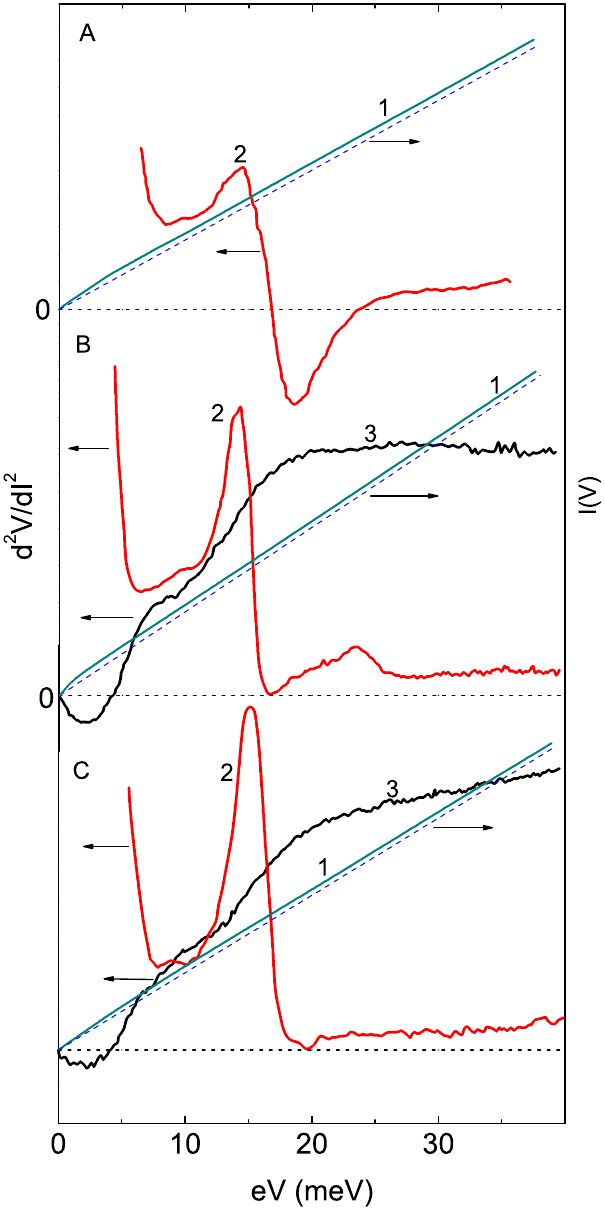}
\caption[]{Volt-ampere characteristics (curve 1) and their second derivatives (curve 2) for the dirty superconducting point-contacts, $T = 4.2-4.5\ K$, $H=0$. The contact resistances $a, b$, and $c$ are equal to 194, 17.5 and $29.2\ \Omega $, respectively. Curve 3 represents the second derivative in the normal state for the junctions b and c ($T=10~K$). The modulation voltages are equal to 0.768 and 2.16 $mV$ for curves 2 and 3 in Fig 2(b), and 0.761 and 2.26~$mV$ in Fig 2(c), respectively.}
\label{Fig2}
\end{figure}
\begin{figure}[]
\includegraphics[width=8cm,angle=0]{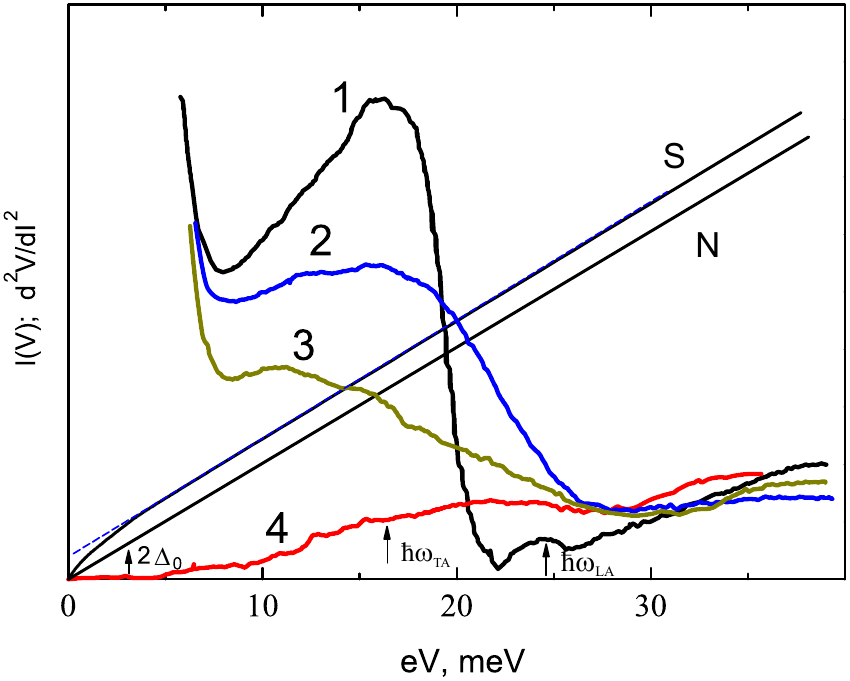}
\caption[]{Point-contact spectra of the same niobium junction taken at different temperatures in zero magnetic field. Curves 1, 2, 3, and 4 correspond to the temperatures 4.4, 6.5, 8.65, and 10.6 $K$. I-V curves in superconducting ($S$) and normal ($N$) states were taken at 7.8 and 11.2 $K$, respectively ${{R}_{0}}=30\ \Omega $. Niobium electrodes are covered by aluminium oxide.}
\label{Fig3}
\end{figure}

The phonon peculiarities in the second derivative curves taken in the $N$-state would be seen only for the pure high ohmic contacts ($R\approx 100\ \Omega $).  Our best PC spectrum, ${{V}_{2}}\left( eV \right)\propto {{d}^{2}}V/d{{I}^{2}}$, is represented by curve 1 in Fig \ref{Fig1}. Since the calculations of ${{g}_{pc}}(eV)$  and ${{\lambda }_{pc}}$  are usually earned out in the frames of the free electron model \cite{Yanson2}, which has not yet been proven to be correct for the description of PC characteristics of transition metals, there is no possibility of making a conclusion about a correspondence of our best contacts to the ballistic regime. But the near equality between $\lambda _{pc}^{\max }$  and ${{\lambda }_{tr}}$  enables us to believe the contacts mentioned above being pure enough Despite the close agreement between ${{\lambda }_{pc}}$ and ${{\lambda }_{tr}}$ for the best contacts the majority of specimens having phonon maxima in their spectra possess much lower values of ${{\lambda }_{pc}}$  which may drop down to 0.01. One would think that the decrease of $g_{pc}$ and ${{\lambda }_{pc}}$  could be due to the dirtiness of the con tact region only, leading to the proportionality between the PC spectrum intensity and ${{l}_{i}}/d$  ratio ($l_i$ and $d$ being the elastic scattering m.f.p. and contact diameter, correspondingly). However, this is not the single cause for $Nb$ since the metallic low order $Nb$ oxides may shunt the contact conductivity. The oxide cannot be avoided because it plays an important supporting role making all the construction rigid enough. Unfortunately the method of oxidation used in this work cannot exclude the low type oxides. As a consequence the calculated contact diameters are greatly overestimated with respect to the true values. This explains why there is no heating in our relatively low ohmic contacts, having either no phonon peaks at all, or only very weak traces of those (see, for example, the spectrum 4 in Fig \ref{Fig3}). The small value of I-V curve nonlinearity (corresponding to about several percentage change of differential conductivity) gives one more evidence on the absence of heating.
The majority of PC spectra corresponding to the dirty but not thermal limit reveals no phonon peculiarities in the normal state at all These spectra have the appearance of curve 3 in Fig \ref{Fig2}(b) and curve 4 in Fig \ref{Fig3}. However, after the transition to superconducting state the strong singularities located at phonon energies appear The shape of the singularities varies from sample to sample but we can roughly separate three types of shapes represented by the curves 2 in Fig \ref{Fig2}(a), (b), and (c), respectively.

One of the limiting forms corresponds to the maximum in differential resistance and has the shape of curve 2 in Fig \ref{Fig2}(a). It should be noted that the same shape is also peculiar to the thermal singularities (the last being noticeably narrower as a rule) which connected with the essential decrease of the excess current. Unlike the singularities which we concentrate on in this paper, the energy positions of the thermal singularities depend on the ambient temperature as well as on the degree of overheating of the sample. The intensity of the thermal singularities was several times higher than that of nonthermal ones. The thermal singularities were probably observed in \cite{Khotkevich}, yet we cannot exclude the possibility that their origin was due not to simple over-heating but to nonequilibrium destroyment of superconductivity instead. We emphasize once more that the overheating in the contacts selected m the present work is negligible because not only the nonlinearity of the I-V curve in normal state is small but also the excess current in superconducting state is constant (within a precision of several per cents) in the whole energy range studied (see I-V curves in Figs. \ref{Fig2} and \ref{Fig3}).

Another limiting case corresponds to the shape being similar to the shape of phonon peaks in the normal state for pure contacts, although the latter definitely were not pure [see curve 2 in Fig. \ref{Fig2}(c)].

The majority of nonthermal PC spectra in the superconducting state has the appearance which was intermediate between two limiting shapes described above [see curve 2 in Fig \ref{Fig2}(b)]. The position of the main maximum may be shifted on 1-2 $meV$ from the position of the transverse phonon density of states peak at 16 $meV$. Some of the spectra reveal the second phonon singularity at 23-25 $meV$ which is due to the longitudinal phonons density of states peak [see curve 2 in Fig. \ref{Fig2}(b)]. In some spectra the soft phonon mode at $eV\sim10\ meV$ is revealed in the normal state as a shoulder [see curve 3 in Fig \ref{Fig2}(b) and (c)].

Figure \ref{Fig3} shows the second derivatives of the I-V characteristics taken at temperatures both higher and lower than $T_c$ of niobium. While approaching $T_c$  from below the spectral singularity at $eV=\hbar {{\omega }_{TA}}$  becomes wider and its intensity falls down but its position at the $eV$-axis does not move towards the smaller values as one would expect if the effect observed were due to heating or to approaching the critical current density in superconducting banks. Furthermore, it is evident that the nonlinearity observed in the superconducting state is completely due to the energy dependence of the excess current and is not connected with the nonlinearity in the normal state. The latter does not exceed 1 -2\%, while for the same contacts being in the superconducting state it grows up to 4-6\% due to the energy dependence of the excess current.

Comparing our experimental results with those of the theoretical works discussing the EPI in superconducting point-contacts one should take into account that in those works the pure contacts are considered while ours are dirty. According to \cite{Khlus1} the EPI in pure superconducting contacts leads to the nonlinearity which amounts only a small fraction (of the order of $\Delta /eV$) of the nonlinearity due to the nonohmic behavior of the I-V curves in the normal state. The latter yields the direct proportionality between the second derivative and PC EPI spectral function \cite{Yanson1}. In several cases we managed to record the $V_{2}(eV)$ characteristics both in $N$- and $S$-states for relatively pure contacts. Qualitatively these spectra appear to be the same as expected from the theory \cite{Khlus1} but the transition from $S$- to $N$-state was accompanied by some changes of contact resistance which prevented us making a definite conclusion.

According to the pure contact theory \cite{Khlus2} the EPI nonlinearity of the I-V curves in the $S$-state should appear as a maximum either on the first or on the second derivative providing the width of the phonon line being less or more than the energy gap, respectively. In our experiments with the dirty contacts the intermediate shapes were the most frequent although the extreme situations were also seen in many cases.

Before making some suggestions about the possible origin of the phenomena found let us roughly estimate the parameters $d,\ {{l}_{i}},\ {{l}_{\varepsilon }}$ and $\xi \left( 0 \right)$ for point-contacts studied. The contact diameter estimated from the resistance usually amounts to several tens of Angstroms and is independent on either pure or dirty limits taken from calculations providing the purity of the surface layer of metal electrodes is the same as the purity of thin films deposited by electron beam evaporation in high vacuum. For these conditions $l_i$ has the same order of magnitude as $d$. As to the energy relaxation mean free path, ${{l}_{\varepsilon }}$, it is known \cite{Shen} not to exceed $\sim100\ \AA$ at Debye energies for transition metals. Only the coherence length appears to be somewhat longer $\xi (0)=0.8\sqrt{{{\xi }_{0}}\cdot {{l}_{i}}}$ (${{\xi }_{0}}=\text{ }400\text{ }\AA$  for pure $Nb$ \cite{Finnemore}). Thus the relation between the contact parameters is as follows:

$\xi (0)\gtrsim d\sim {{\Lambda }_{\varepsilon }}$\ \
where ${{\Lambda }_{\varepsilon }}=\sqrt{{{l}_{i}}\cdot {{l}_{\varepsilon }}}$.

At these conditions one should expect the PCS to be hindered as it is observed in the experiment. However the heating is not seen in our experiments that may be due to the phonon mean free path being larger than the contact region consisted of dirty superconducting material.

Taking into consideration that the subharmonic gap structure at $eV=\Delta$ and $2\Delta$ is seen in the I-V curves of the majority of the junctions, and the excess current is close to the theoretical value for the dirty $S-c-S$ contacts, one may suppose that just this type of junction holds in our case. However, the low values of the energy gaps ($\Delta=1.2\div1.5 \ meV$ instead of $\Delta=1.55$ for pure $Nb$) and the absence of the superconducting current evidence that the superconductivity is strongly suppressed in the centre of the contact.

Concerning the reason which allows the PCS to be possible in the dirty superconducting contacts one may speculate as follows. In general, the PCS of phonons is due to the presence of two groups of electrons having maximum energy difference equal to $eV$ in the same metallic region. In normal state for dirty contacts ($d\gtrsim\Lambda_\varepsilon$) this condition is not satisfied and no phonon spectral lines can be seen. However, in the superconducting state it may appear again that inside the contact region with the characteristic dimensions of the order of $\Lambda_\varepsilon$ there are two groups of electrons differing in energy by $eV$. These electrons arise via the decay of the Cooper pairs. At the left and right sides of the contact the chemical potentials of pairs are constant and differ by $eV$ (with respect to a single electron) despite the presence of the electric field and the continuous change of the chemical potential of quasi- particles at the distance of the order of $d$. Inelastic single electron transitions between the pair levels may be responsible for the phonon singularities observed in I-V characteristics of the dirty superconducting contacts.


\begin{thebibliography}{}
\bibitem{Yanson1} I.K. Yanson, JEPT 66, 1035 (1974).
\bibitem{Kulik} I.O. Kulik, A.N. Omelyanchuk and R.I. Shekhter, Fizika Nizkikh Temperatur 3, 1267 (1977).
\bibitem{Yanson2} I.K. Yanson, \href{http://fntr.ilt.kharkov.ua/fnt/pdf/9/9-7/f09-0676r.pdf}{Fiz. Nizk. Temp.} 9, 676 (1983) [Sov. J. Low Temp. Phys. 9, (1983)].
\bibitem{Khotkevich} A.V. Khotkevich and I.K. Yanson, \href{http://fntr.ilt.kharkov.ua/fnt/pdf/7/7-6/f07-0727r.pdf}{Fiz. Nizk. Temp.} 7, 727 (1981) [Sov. J. Low Temp. Phys. 7, 354 (1981)].
\bibitem{Khlus1} V.A. Khlus and A.N. Omel'yanchuk, \href{http://fntr.ilt.kharkov.ua/fnt/pdf/9/9-4/f09-0373r.pdf}{Fiz. Nizk. Temp.} 9, 373 (1983) [Sov. J. Low Temp. Phys. 9, 189 (1983)].
\bibitem{Divin}Yu.Ya. Divin and F.Ya. Nad', Fizika Nizkikh Temperatur 4, 1105 (1978).
\bibitem{Wolf} E.L. Wolf and J. Zasadsinski, Inst. Phys. Conf. Ser. No. 39, (1968), Chap. 8, pp. 667-670.
\bibitem{Goodman}B.B. Goodman and G. Kuhn, \href{http://www.slac.stanford.edu/cgi-wrap/getdoc/slac-trans-0139.pdf}{J.Phys} (Paris) 29, 240 (1968).
\bibitem{Robinson} B. Robinson, T.H. Geballe, and J. M. Rowell, "Tunneling study of niobium using aluminum-aluminum oxide-niobium junctions" in: \href{http://link.springer.com/chapter/10.1007%2F978-1-4615-8795-8_21} {Superconductivity in d- and f-band Metals}, D. H. Douglass (ed.), Plenum Press, New York (1976), pp. 381-386.

\bibitem{Khlus2} V.A. Khlus, \href{http://fntr.ilt.kharkov.ua/fnt/pdf/9/9-9/f09-0985r.pdf}{Fiz. Nizk. Temp.} 9, 985 (1983) [Sov. J. Low Temp. Phys. 9, (1983)].

\bibitem{Shen}L.J.L. Shen, \href{https://www.researchgate.net/publication/234851839_Superconductivity_of_Tantalum_Niobium_and_Lanthanum_Studied_by_Electron_Tunneling_Problems_of_Surface_Contamination}{Superconductivity in d- and f-band metals }(Edited by D H Douglass) pp. 31-44, Plenum Press, New York (1972).
\bibitem{Finnemore}D.K. Finnemore, T.F. Stromberg and C.A. Swenson, \href{http://journals.aps.org/pr/abstract/10.1103/PhysRev.149.231}{Phys. Rev.} 149, 231 (1966).
	
	
\end{thebibliography}
\end{document}